\documentstyle [12pt] {article}
\makeatletter
\ifcase \@ptsize
\or % mods for 11 pt
\or % mods for 12 pt
\fi
\advance\textheight by \topskip

\ifcase \@ptsize
\or % mods for 11 pt
\or % mods for 12 pt
\fi

\def\@citex[#1]#2{%
\if@filesw \immediate \write \@auxout {\string \citation {#2}}\fi
\@tempcntb\m@ne \let\@h@ld\relax \def\@citea{}%
\@cite{%
  \@for \@citeb:=#2\do {%
    \@ifundefined {b@\@citeb}%
      {\@h@ld\@citea\@tempcntb\m@ne{\bf ?}%
      \@warning {Citation `\@citeb ' on page \thepage \space undefined}}%
      {\@tempcnta\@tempcntb \advance\@tempcnta\@ne%
      \@tempcntb\number\csname b@\@citeb \endcsname \relax%
      \ifnum\@tempcnta=\@tempcntb %   Number follows previous--hold on to it
        \ifx\@h@ld\relax%
          \edef \@h@ld{\@citea\csname b@\@citeb\endcsname}%
        \else%
          \edef\@h@ld{\ifmmode{-}\else--\fi\csname b@\@citeb\endcsname}%
        \fi%
      \else%   %  non-successor--dump what's held and do this one
        \@h@ld\@citea\csname b@\@citeb \endcsname%
        \let\@h@ld\relax%
      \fi}%
    \def\@citea{,\penalty\@highpenalty\,}%
  }\@h@ld%
}{#1}}

\makeatother
\begin{document}
\hfuzz=100pt
%\rightmargin -2.75cm
%\textheight 23.0cm
%\topmargin -0.5in
%\baselineskip 16pt
%\parskip 18pt
%\parindent 30pt
%\def\mc{\,\raise -2.truept\hbox{\rlap{\hbox{$\sim$}}\raise5.truept
%\hbox{$<$}\ }}
%\def\Mc{\,\raise -2.truept\hbox{\rlap{\hbox{$\sim$}}\raise5.truept
%\hbox{$>$}\ }}%
%
%
%%%%%%%%%%%%%%%%%%%%%%%%%%%%%%%%%%%%%%%%%%%%%%%%%%%%%%%%%%%%%%%
%
%    List of  the     commands
%%%%%%%%%%%%%%%%%%%%%%%%%%%%%%%%%%%%%%%%%%%%%%%%%%%%%%%%%%%%%%%
%
\newcommand{\be}{\begin{equation}}
\newcommand{\ee}{\end{equation}}
\newcommand{\bea}{\begin{eqnarray}}
\newcommand{\eea}{\end{eqnarray}}
\begin{titlepage}
\makeatletter
\def \thefootnote {\fnsymbol {footnote}} \def \@makefnmark {
\hbox to 0pt{$^{\@thefnmark }$\hss }}
\makeatother
\begin{flushright}
BONN-HE-92-38\\
December, 1992
\end{flushright}
\vspace{2cm}
\begin{center}
{ \large \bf Arbitrary Spacetimes from the ${\bf{SL(2,R)/U(1)}}$ Coset Model}\\
\vspace{2cm}
{\large\bf Noureddine Mohammedi} \footnote
{Work supported by the Alexander von Humboldt-Stiftung.}
\footnote{e-mail: nouri@avzw02.physik.uni-bonn.de}
\\
\vspace{.5cm}
\large Physikalisches Institut\\
der Universit\"at Bonn\\
Nussallee 12\\ D-5300 Bonn 1, Germany\\

\baselineskip 18pt
\vspace{.2in}
\vspace{1cm}
{\large\bf Abstract}
\end{center}
We show that the gauged $SL(2,R)$ WZWN model yields arbitrary
spacetimes in two
dimensions. The $c\,=\,1$ matter coupled to gravity and the black hole
singularity are just
two particular cases in these spacetimes.
\\

\setcounter {footnote}{0}
\end{titlepage}
\baselineskip 20pt
%

%\setcounter{chapter}{1}
%\setcounter{section}{1}
%\setcounter{subsection}{1}
%\section{Introduction}
%\setcounter{equation}{0}

\par
Since the discovery of the existence of a two dimensional black hole
in the coset model $SL(2,R)/U(1)$ [1] and as a solution to the string beta
functions [2,3], many efforts have been devoted to this subject.
In particular, attempts have been directed towards understanding the
relationship
between the black hole and the $c\,=\,1$ matter field coupled to
two dimensional gravity [4-10]. All the evidences point to the fact that
these two theories might have the same origin.
\par
In this note, we show that by gauging the $U(1)$ isometry subgroup of the
$SL(2,R)$ WZWN model one obtains an arbitrary geometry for the two
dimensional target-space. The $c\,=\,1$ matter coupled to gravity and the
black hole are just two particular cases in this arbitrary geometry.
\par
This arbitrariness in the geometry arises from the mathematical formalism
of gauging an isometry subgroup of a general non-linear sigma model [11,12].
This formalism was also recently used to identify the perturbations,
by $(1,1)$ conformal operators, of the black hole found in the
$SL(2,R)/U(1)$ coset model [13].
\par
Our starting point is the ungauged $SL(2,R)$ WZWN action
\bea
I(g)&=&{k\over 8\pi}\int_\Sigma d^2x\sqrt{\gamma}\gamma^{\mu\nu}Tr\left[\left(
g^{-1}\partial_\mu g\right)\left(g^{-1}\partial_\nu g\right)\right]\nonumber\\
&+&{k\over 12\pi}\int_B d^3y\epsilon^{\mu\nu\rho}Tr\left[\left(
g^{-1}\partial_\mu g\right)\left(g^{-1}\partial_\nu g\right)
\left(g^{-1}\partial_\rho g\right)\right]\,\,\,.
\eea
Here $B$ is a three dimensional manifold whose boundary is $\Sigma$ and $Tr$ is
the trace in the two dimensional representation of $SL(2,R)$. Let us
parametrize the $SL(2,R)$ group manifold by
\be
g=\left( \begin{array}{cc}
a&u\\
-v&b
\end{array} \right)\,\,\,,\,\,\,ab+uv=1\,\,\,.
\ee
In this parametrization, the above action yields
\be
I(\phi)={k\over 4\pi}\int {\rm {d}}^2x\left(\sqrt {\gamma}\gamma^{\mu\nu}G_{ij}
+\epsilon^{\mu\nu}B_{ij}\right)\partial_\mu\phi^i\partial_\nu\phi^j\,\,\,,
\ee
where the target-space metrice $G_{ij}$ and the antisymmetric tensor
$B_{ij}$ are
\be
G_{ij}=\left(\begin{array}{ccc}
{1\over a^2}(1-uv)&{1\over 2}{v\over a}&{1\over 2}{u\over a}\\
{1\over 2}{v\over a}&0&-{1\over 2}\\
{1\over 2}{u\over a}&-{1\over 2}&0
\end{array}\right )\,\,,\,\,
B_{ij}=\left(\begin{array}{ccc}
0&0&0\\
0&0&-\ln a\\
0&\ln a&0
\end{array}\right )\,\,\,.
\ee
Here $\phi^1\,=\,a$, $\phi^2\,=\,u$, $\phi^3\,=\,v$.
\par
A general non-linear sigma model as given in (3) possesses a global $U(1)$
isometry
symmetry given by
\be
\delta\phi^i=\varepsilon K^i(\phi)\,\,\,,
\ee
provided that $K^i$ is a Killing vector of the metric $G_{ij}$ and the
antisymmetric tensor $B_{ij}$ satisfies
\be
\partial_l B_{ij}K^l +B_{lj}\partial_iK^l+B_{il}\partial_jK^l=
\nabla_iL_j -\nabla_jL_i\,\,\,,
\ee
for some target-space vector $L_i$ [14].
\par
This global $U(1)$ symmetry can be made local by introducing a $U(1)$ gauge
field $A_\mu$ transforming as
\be
\delta A_\mu=-\partial_\mu\varepsilon\,\,\,.
\ee
The most general action involving the gauge field $A_\mu$ is
written as [11,12]
\be
I_{gauged}={k\over 4\pi}\int d^2x\{\sqrt {\gamma}\gamma^{\mu\nu}G_{ij}
D_\mu\phi^iD_\nu\phi^j+\epsilon^{\mu\nu}B_{ij}
\partial_\mu\phi^i\partial_\nu\phi^j
-2\epsilon^{\mu\nu}C_iA_\mu\partial_\nu\phi^i\}\,\,\,,
\ee
where
\be
D_\mu\phi^i= \partial_\mu\phi^i+A_\mu K^i
\ee
and the target-space function $C_i(\phi)$ is given by
\be
C_i=B_{ij}K^j+L_i\,\,\,.
\ee
Local gauge invariance implies then the following equations [11,12]
\bea
&\partial_jC_iK^j+C_j\partial_iK^j=0&\\
&L_iK^i=0&\,\,\,.
\eea
\par
Eliminating the gauge fields from (8) results in a new non-linear sigma
model of the form (3) with a new metric  $\widehat{G}_{ij}$ and a new
antisymmetric tensor $\widehat{B}_{ij}$ given by
\bea
\widehat{G}_{ij}&=& G_{ij}-{1\over M}\left(G_{ik}G_{jl}K^kK^l - C_iC_j\right)
\nonumber\\
\widehat{B}_{ij}&=& B_{ij}+{1\over M}\left(G_{ik}C_jK^k -G_{jk}C_iK^k
\right)\,\,\,,
\eea
where
\be
M=G_{ij}K^iK^j\,\,\,.
\ee
Notice that the new metric $\widehat{G}_{ij}$ would exhibit an explicit
singularity
if $M$ has zeros. This is so if the old metric $G_{ij}$ is not
positive definite as it is in the case when the non-linear sigma model is
defined on a  non-compact group manifold. Using equation (13) we find
\be
\widehat{G}_{ij}K^j=0\,\,\,.
\ee
Therefore due to these null eigenvectors, the metric $\widehat{G}_{ij}$ cannot
be inverted and we cannot analyse the singularities of the gauged
non-linear sigma model. To overcome this difficulty,  a gauge fixing term
must be introduced.
\par
We would like now to apply this analyses to the non-linear sigma model of
the $SL(2,R)$ WZWN action. For this, we would like to gauge
the non-compact one parameter symmetry group generated by
\be
\delta g=\varepsilon\left\{\left(\begin{array}{cc}
1&0\\0&-1\end{array}\right)g+g\left(\begin{array}{cc}
1&0\\0&-1\end{array}\right)\right\}\,\,\,.
\ee
This transformation is of the form (5) where the corresponding Killing vectors
are
given by
\be
K^1=2a\,\,,\,\,K^2=0\,\,,\,\,K^3=0\,\,\,\,.
\ee
We deduce from equations (4) and (10) that
\be
C_1=L_1\,\,,\,\,C_2=L_2\,\,,\,\,C_3=L_3\,\,\,.
\ee
Furthermore, equation (12) leads to
\be
C_1=L_1=0
\ee
while equation (11) is solved by
\be
C_2=f(u,v)\,\,\,,\,\,\,C_3=h(u,v)\,\,\,,
\ee
where $f(u,v)$ and $h(u,v)$ are two arbitrary functions.
These two functions are, however, not independent. They are related
by the defining equation for $L_i$. Indeed, equation (6) leads to
\be
2=\partial_vf(u,v)-\partial_uh(u,v)\,\,\,.
\ee
This differential equation has the following general solution
\bea
f(u,v)&=&v-\partial_uX(u,v)\nonumber\\
h(u,v)&=&-u-\partial_vX(u,v)\,\,\,,
\eea
where $X(u,v)$ is an arbitrary function. Therefore the quantities $C_1$,
$C_2$ and $C_3$ have been completely determined.
Consequently, the non-vanishing components of the new metric $\widehat
{G}_{ij}$
are listed below
\bea
\widehat{G}_{22}&=&-{1\over 4(1-uv)}\left[2v\left(\partial_uX\right)
-\left(\partial_uX
\right)^2\right]\nonumber\\
\widehat{G}_{23}&=&-{1\over 4(1-uv)}\left[2-u\left(\partial_uX\right)
+v\left(\partial_vX\right)-
\left(\partial_uX\right)\left(\partial_vX\right)\right]\nonumber\\
\widehat{G}_{33}&=&-{1\over 4(1-uv)}\left[-2u\left(\partial_vX\right)
-\left(\partial_vX\right)^2\right]\,\,\,.
\eea
\par
The scalar curvature corresponding to this metric is found to be
\bea
R&=&{A\over B}\nonumber\\
A&=&
-64 +96 \left[v \left(\partial_uX\right) \left(\partial_vX\right)^2-u
\left(\partial_vX\right) \left(\partial_uX\right)^2\right]\nonumber\\
&+&32\left[ 2+5 u v-u^2 v^2\right] \left(\partial_uX\right)
\left(\partial_vX\right)\nonumber\\
&-&8\left[ 4-2 u v+u^2 v^2\right] \left(\partial_uX\right)^2
\left(\partial_vX\right)^2\nonumber\\
&+&8 \left(3-u v\right) \left[u^2 \left(\partial_uX\right)^3
\left(\partial_vX\right)+v^2 \left(\partial_vX\right)^3
\left(\partial_uX\right)\right]\nonumber\\
&+&32\left[ 1-3 u v+3 u^2 v^2-u^3 v^3\right]\left[
\left(\partial_u^2X\right) \left(\partial_v^2X\right)
-\left(\partial_v\partial_uX\right)
^2\right]\nonumber\\
&+&4 u \left(\partial_uX\right)\left[ 32-24 u \left(\partial_uX
\right)+8 u^2 \left(\partial_uX\right)^2-u^3
\left(\partial_uX\right)^3\right]\nonumber\\
&-&4 v \left(\partial_vX\right)\left[ 32+24 v
\left(\partial_vX\right)+8 v^2 \left(\partial_vX\right)^2+v^3
\left(\partial_vX\right)^3\right]\nonumber\\
&+&\left(1-u v\right)^2 \left[32 \left(\partial_uX\right)^2
\left(\partial_v^2X\right)-32
\left(\partial_vX\right)^2 \left(\partial_u^2X\right)\right]\nonumber\\
&+&\left(1-u v\right)^2 \left[-8 u \left(\partial_uX\right)^3
\left(\partial_v^2X\right)-8 v \left(\partial_vX\right)^3
\left(\partial_u^2X\right)\right]\nonumber\\
&+&\left(1-u v\right)^2 \left[-32 u \left(\partial_u^2X\right)
\left(\partial_vX\right)-32 v \left(\partial_v^2X\right)
\left(\partial_uX\right)\right]\nonumber\\
&+&\left(1-u v\right)^2 \left[-8 u \left(\partial_u^2X\right)
 \left(\partial_uX\right) \left(\partial_vX\right)^2-8 v
\left(\partial_v^2X\right)\left(\partial_vX\right)
\left(\partial_uX\right)^2\right]\nonumber\\
&+&\left(1-u v\right)^2\left[ 16 u \left(\partial_v\partial_uX\right)
\left(\partial_uX\right)^2 \left(\partial_vX\right)+16 v
\left(\partial_v\partial_uX\right)\left(\partial_vX\right)^2
\left(\partial_uX\right)\right]\nonumber\\
B&=&
-\left(1-uv\right)
\left[4-4u\left(\partial_uX\right)+4v\left(\partial_vX\right)
+u^2\left(\partial_uX\right)^2+v^2\left(\partial_vX\right)^2\right.
\nonumber\\
&-&\left.2\left(2-uv\right)\left(\partial_uX\right)
\left(\partial_vX\right)\right]^2\,\,\,.
\eea
\par
Therefore, due to the arbitrariness of the function $X(u,v)$, we can
obtain {\it{any geometry}} we like from gauging the $U(1)$ isometry
subgroup of the $SL(2,R)$ WZWN model. In particular if we choose
\be
X(u,v)=constant
\ee
then the metric $\widehat{G}_{ij}$ exhibits the black hole geometry
found in [1].
On the other hand, the $c\,=\,1$ matter field coupled to two
dimensional gravity is realized by all the functions $X(u,v)$ which
are solutions to the differential equation
\be
A=0\,\,\,.
\ee
\par
This arbitrariness in the geometry is better understood in the
language  of the $SL(2,R)$ group manifold.
It turns out that the metric $\widehat{G}_{ij}$ in (23)
arises, upon eliminating the gauge field,
from the following action
\bea
I(g,A)&=&I(g)+{k\over 4\pi}\int d^2x \sqrt\gamma\gamma^{\mu\nu}
Tr\left\{A_\mu\left(\begin{array}{cc}1&0\\0&-1\end{array}\right)g^{-1}\partial_\nu g
+A_\mu\left(\begin{array}{cc}1&0\\0&-1\end{array}\right)\partial_\nu
gg^{-1}\right.
\nonumber\\
&+&\left.A_\mu A_\nu\left[\left(\begin{array}{cc}1&0\\0&-1\end{array}\right)
\left(\begin{array}{cc}1&0\\0&-1\end{array}\right)
+\left(\begin{array}{cc}1&0\\0&-1\end{array}\right)g
\left(\begin{array}{cc}1&0\\0&-1\end{array}\right)g^{-1}\right]\right\}\nonumber\\
&+&{k\over 4\pi}\int d^2x \epsilon^{\mu\nu}
Tr\left[A_\mu\left(\begin{array}{cc}1&0\\0&-1\end{array}\right)g^{-1}\partial_\nu g
-A_\mu\left(\begin{array}{cc}1&0\\0&-1\end{array}\right)\partial_\nu
gg^{-1}\right]
\nonumber\\
&+&{k\over 4\pi}\int d^2x \epsilon^{\mu\nu}
Tr\left[\partial_\mu A_\nu\left(\begin{array}{cc}1&0\\0&-1\end{array}\right)
\left(\begin{array}{cc}X&0\\0&-X\end{array}\right)\right]
\,\,\,.
\eea
\par
Without the last term, this action is what is commonly written down for the
$SL(2,R)/U(1)$ WZWN model [15,16] when gauging the $U(1)$ subgroup as
given in (16).
The last term, however, is  a topological term
\be
{k\over 4\pi}\int d^2x \epsilon^{\mu\nu}F_{\mu\nu}X\left(u,v\right)
\ee
whose presence is required by the mathematical structure arising from
gauging the $U(1)$ isometry of a general non-linera sigma model. Therefore,
there is no reason for ignoring such a term when dealing with  gauged WZWN
models.
Furthermore, this is the same term which was interpreted in ref.[13] as
generating
perturbations, by $(1,1)$ conformal operators, of the $SL(2,R)/U(1)$ black
hole.

\vspace{0.5cm}

\paragraph{Acknowledgements:} I would like to thank W. Nahm for showing his
interest in the subject. The
financial support  from the Alexander von
Humboldt-Stiftung is also hereby acknowledged.

\end{document}